\documentclass[doublecol,final,a4paper]{epl2_my}
\usepackage{graphicx}
\usepackage{amsmath}
\usepackage{amssymb}

\begin{document}

\title{Entanglement in Theory Space}
\author{Masahito Yamazaki}
\institute{Princeton Center for Theoretical Science,
 Princeton University, Princeton NJ 08540, USA}

\abstract{
We propose a new concept of entanglement for quantum systems: 
{\it entanglement in theory space}.
This is defined by decomposing a theory into two by an un-gauging procedure.
We provide two examples where this newly-introduced entanglement is
 closely related to conventional geometric entropies: deconstruction
 and AGT-type correspondence.
}

\received{24 April 2013}
\acceptedinfinalform{9 July 2013}
\onlinepub{xxxx August 2013}


\maketitle

{\bf Introduction.}---
Entanglement entropy is an indispensable measure for
intrinsically quantum properties of quantum systems, and 
plays crucial roles in a number of different disciplines,
such as quantum information and computation, many-body systems, 
quantum field theories and black hole physics 
(see {\it e.g.} Refs.~\cite{reviews,CC,RTreview} for reviews).

The goal of this Letter is to 
introduce a new concept of entanglement.
As we will review momentarily the conventional definition of 
entanglement entropy involves the 
division of a spatial region into two.
By contrast our entanglement entropy is defined by
decomposing a gauge theory 
into two by an 
un-gauging of part of the gauge symmetry.
Since the decomposition here refers not
to geometric spatial regions but to more 
abstract ``space of quantum theories'', 
our entanglement entropy will be 
called {\it entanglement in theory space},
or {\it theory-space entanglement};
for definiteness the conventional concept of entanglement
will be hereafter called {\it geometric entanglement}.

While the concept of theory-space entanglement is 
rather unexplored and deserves further study,
we point out that there are some examples where 
theory-space entanglement entropy is closely related to, or even 
equal to, the geometric entanglement entropy.

{\bf Geometric Entanglement.}---
Let us first briefly recall the more conventional version,
{\it i.e.} the geometric entanglement entropy.

Suppose that we have a quantum mechanical system;
this could either be a discrete lattice system or a continuous field theory.
In the canonical quantization
we obtain a Hilbert space $\mathcal{H}_{\rm tot}$ on a time slice.
Let us divide the spatial regions into a region $A$ and its complement $B$.
The total Hilbert space then factorizes 
into a product of those associated with 
regions $A$ and $B$:
\begin{equation}
\mathcal{H_{\rm tot}}=\mathcal{H}_A\otimes \mathcal{H}_B \ .
\label{Hfactor}
\end{equation}

Now consider the ground state of the total theory
and the associated density matrix $\rho_{\rm tot}$.
We can then define the reduced density matrix by
\begin{equation}
\rho_A=\textrm{Tr}_{\mathcal{H}_B} \rho_{\rm tot} \ ,
\label{def1}
\end{equation}
and the entanglement entropy $S_{\rm ent}$ as the 
von Neumann entropy for $\rho_A$: \footnote{
The entanglement entropies for $A$ and $B$
coincide when $\rho_{\rm tot}$ is constructed from a pure state in the total
Hilbert space $\mathcal{H}_{\rm tot}$.
}
\begin{equation}
S_{\rm ent}=-\textrm{Tr}_{\mathcal{H}_A} \rho_A \log \rho_A \ .
\label{def2}
\end{equation}

{\bf (Un-)Gauging.}---
In the definition above of the geometric entanglement entropy, 
the essential ingredients are that
(i) there is a Hilbert space $\mathcal{H_{\rm tot}}$ and the ground
state density matrix $\rho_{\rm tot}$; (ii) the 
total Hilbert space factorizes as in \eqref{Hfactor}.
We can then define the entanglement  entropy $S_{\rm ent}$
by \eqref{def1}, \eqref{def2}.

While the spatial division gives rise to natural decomposition of the 
Hilbert space, it is not the only possibility.
For example, Ref.~\cite{BMR} proposes entanglement entropy 
in the momentum space.
Our proposal in this Letter 
is more drastic, and relies on the gauging/un-gauging procedure,
which we now explain.

Suppose that we have two theories $\mathcal{T}_A$ and $\mathcal{T}_B$,
with global symmetries $G_A$ and $G_B$, respectively.
Concretely this means that we have two different Lagrangians
$\mathcal{L}_A$ and $\mathcal{L}_B$ with global symmetries
 $G_A$ and $G_B$, respectively.
 We assume the two theories are weakly gauged, i.e.,
 the current $j^{\mu}_{A}$ ($j^{\mu}_{B}$) for the global symmetry 
 could be coupled with the background gauge field $\mathcal{A}^{\mu}_{A}$ ($\mathcal{A}^{\mu}_{B}$)
 by including a term $\int \mathcal{A}^{\mu}_{A} j_{\mu}$ ($\int \mathcal{A}^{\mu}_{B} j_{\mu}$)
 in the Lagrangian $\mathcal{L}_A$ ($\mathcal{L}_B$).
 Note that at this point the gauge fields $\mathcal{A}^{\mu}$
 do not have kinetic terms and hence are not yet dynamical. 

Suppose now that $G_{A}$ and $G_{B}$ contain a common subgroup $G$.
We can then define a new theory $\mathcal{T}_{\rm tot}$ 
by (1) first identifying the $G$-components of the corresponding background gauge fields 
$\mathcal{A}_A$ and $\mathcal{A}_B$ and 
(2) second adding a kinetic term $\frac{1}{g^2} \textrm{Tr} F_{\mu \nu}F^{\mu\nu}$ for the 
$G$-gauge field identified in the first step, where $F_{\mu\nu}$ is the field strength for the gauge field $\mathcal{A}^{\mu}$.

After this gauging, the two theories $\mathcal{T}_A$ and 
$\mathcal{T}_B$ now interact with each other 
through the dynamical
gauge field, and should be regarded as a single interacting theory 
$\mathcal{T}_{\rm tot}$. The coupling constant $g$ determines how strong this 
interaction is.
In the limit $g\to 0$.
the theory $\mathcal{T}_{\rm tot}$ decomposes into 
two decoupled theories $\mathcal{T}_A$
and $\mathcal{T}_B$, each of which are coupled only with the 
non-dynamical background gauge fields (i.e. weakly gauged).
This is known as un-gauging, the opposite of the gauging procedure.

\begin{figure}[htpb]
\centering{\includegraphics[scale=0.3]{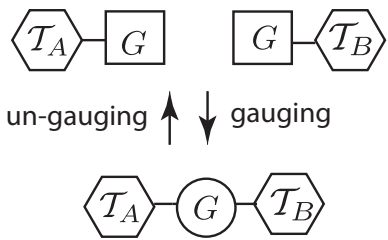}}
\caption{Graphical representation of gauging/un-gauging.
Here hexagons represent the theories, squares global symmetry, and a
 circle a gauge symmetry.
}\label{Tglue}
\end{figure}

This definition of gauging/un-gauging does not really require the 
Lagrangian descriptions, and in fact some of the examples we discuss later are without 
Lagrangians. 
Suppose again that we have two theories $\mathcal{T}_A$ and $\mathcal{T}_B$
with global symmetries $G_A$ and $G_B$, respectively.
We can then gauge the diagonal 
$G$-symmetry inside $G_A \times G_B$ (Fig.~\ref{Tglue}) to define a new theory 
$\mathcal{T}_{\rm tot}$.\footnote{
After gauging the commutant of $G$ inside 
$G_{A,B}$ remains as global symmetries of
$\mathcal{T}_{A,B}$.
} We schematically write this as
\begin{equation}
\mathcal{T_{\rm tot}}=\mathcal{T}_A \cup_G \mathcal{T}_B \ .
\label{Tfactor}
\end{equation}
The details of gauging might differ depending on the
symmetries we wish to preserve. For example, 
when $\mathcal{T}_{A}$ and $\mathcal{T}_B$ have supersymmetry
we can supersymmetrize the gauging procedure \eqref{Tfactor} by
adding superpartners to the coupling $\int A^{\mu} j_{\mu}$.

The gauging procedure described here is 
rather general, and can describe a wide variety of phenomena 
involving gauge fields. For example, 
we can take the subtheories $\mathcal{T}_A$ and $\mathcal{T}_B$
to be the states of atoms, each interacting with the background photons;
the atoms interact with each other only through the long-range interaction
mediated by photons. Another example 
coming from high-energy physics is the 
gauge-mediated supersymmetry breaking \cite{GMSB},
where $\mathcal{T}_A$ is the supersymmetry-breaking sector
(together with messengers),
$\mathcal{T}_B$ is the supersymmetric 
generalizations of standard model,
and the gauge group $G$ is the 
$SU(5)$ gauge group for the grand unified theory.

As these examples show the decomposition \eqref{Tglue}
is a {\it physical} decomposition --- 
it is a choice of the duality frame.
This should be contrasted with the
case of the geometric entanglement entropy,
whose decomposition of spatial regions is often non-physical; 
we can define geometric entanglement entropy for a region $A$ with 
any shape, and we use the choice of $A$ to extract different 
quantities representing entanglement of the theory.

More conceptually,
we can regard the gauging as a 
procedure of constructing more complicated theories 
out of simple ingredients,
and by repeating this procedure we obtain a zoo of 
quantum (field) theories with rich structures.
This viewpoint has been recently explored extensively
in the context of supersymmetric gauge theories,
and we will discuss some of these examples later.

{\bf Theory-Space Entanglement.}---
We can now define the theory-space entanglement. 

Suppose that $\mathcal{T}_{\rm tot}$ is a $D$-dimensional theory
obtained by gauging two $D$-dimensional theories $\mathcal{T}_A$ 
and $\mathcal{T}_B$,
as in \eqref{Tfactor}.
Let us consider the theory $\mathcal{T}_{\rm tot}$ on a $D$-dimensional 
manifold of the form
$\mathbb{R}_t\times \mathcal{S}$, where $\mathcal{S}$ is a compact
$(D-1)$-dimensional manifold 
and $\mathbb{R}_t$ is the time direction \footnote{
It is not necessary to consider Lorentzian signature.
For Euclidean signature the ``time'' is just one of the directions inside
$D$-dimensions.
}. In the canonical quantization
we obtain a Hilbert space $\mathcal{H}_{\rm tot}$ for a fixed time $t$,
and the ground state density matrix $\rho_{\rm tot}$.
By repeating this procedure for $\mathcal{T}_{A}$ ($\mathcal{T}_B$) we also obtain 
$\mathcal{H}_{A}$ ($\mathcal{H}_B$).

The basic idea is now clear: since we have $\mathcal{H}_{\rm tot}$,  
$\mathcal{H}_{A}$, $\mathcal{H}_B$ and $\rho_{\rm tot}$,
we can define the entanglement entropy by using the same formulas 
\eqref{def1}, \eqref{def2}.

There is one important subtlety, however;
the factorization of the Hilbert space 
\eqref{Hfactor} does {\it not} hold, and we only 
have an embedding
\begin{equation}
\iota: \mathcal{H}_{\rm tot} \hookrightarrow \mathcal{H}_A\otimes \mathcal{H}_B \ .
\label{Hembed}
\end{equation}
The reason is that 
the states $|\psi_{A}\rangle$ in $\mathcal{H}_{A}$ ($|\psi_{B}\rangle$ in $\mathcal{H}_{B}$) in general 
is charged non-trivially under the global symmetry $G$ before un-gauging,
but then the
product state $|\psi_A\rangle \otimes |\psi_B\rangle$ does not make sense
as a state of $\mathcal{H}_{\rm tot}$ since it is charged under $G$,
which is now promoted to a gauge symmetry in $\mathcal{T}_{\rm tot}$ after gauging.
Nevertheless we can defined the embedding \eqref{Hembed}
by incorporating the degrees of freedom for the gauge group $G$
(and their superpartners) in the definition of $\mathcal{H}_{A,B}$, 
thus effectively doubling the degrees of freedom of $G$.
To emphasize this some readers might prefer the notation $\mathcal{H}_{A+G}, \mathcal{H}_{B+G}$.

The non-factorization however is not really a problem, and a small modification
saves the definition.
The embedding $\iota$ induces the embedding of the 
ground state density matrix $\rho_{\rm tot}=|\psi_0 \rangle \langle \psi_0|$:
\begin{equation}
\iota^* (\rho_{\rm tot}):=\iota \big(|\psi_0 \rangle\big) \iota \big(\langle \psi_0|\big)
 \ .
\end{equation}
Note by definition $\iota$ maps a pure state into a pure state.
We modify the equation \eqref{def1} by
\begin{equation}
\rho_A=\textrm{Tr}_{\mathcal{H}_B} \iota^*(\rho_{\rm tot}) \ ,
\label{def3}
\end{equation}
and can define the theory-space entanglement by the same equation 
\eqref{def2}. By the Schmidt decomposition it follows immediately  
that the answer does not change when we exchange the roles of 
$A$ and $B$. This concludes our definition of theory-space
entanglement. \footnote{We can generalize the definition 
to the case where we gauge the 
diagonal global symmetry for a set of theories $\mathcal{T}_{A_i}$, 
each with a global symmetry $G$.
In the graphical representation of Fig.~\ref{Tglue} this will be
a multi-valent vertex.
}

The theory-space entanglement defined here depends non-trivially
on the gauge coupling constant $g$ for the dynamical gauge field,
as well as on the choice of the ground state wavefunction.
The latter choice will be crucial for the supersymmetric examples
discussed in the latter part of this Letter.

The definition of the theory-space entanglement requires not only
the 
theory $\mathcal{T}$ itself, but also 
the choice of the decomposition \eqref{Tfactor}
and the compactification manifold
$\mathcal{S}$. Neither of these choices is unique.
 
The choice of the decomposition \eqref{Tfactor}
comes in since (as discussed above) we need a physical 
decomposition of the Hilbert space.
This is closely related to the issue of duality;
the theory $\mathcal{T}_{\rm tot}$, defined in \eqref{Tfactor},
in general could have a different decomposition
\begin{equation}
\mathcal{T_{\rm tot}}=\mathcal{T}_{A'} \cup_{G'} \mathcal{T}_{B'} \ ,
\label{Tfactor2}
\end{equation}
and $G'$ can be rather different from the gauge group $G$ in another frame;
gauge symmetry is by definition a redundancy for describing physics,
and there is no unique way to associate a unique 
gauge symmetry for a given physical system.\footnote{A 
prototypical examples for this is the Seiberg duality \cite{Seiberg} for 4d $\mathcal{N}=1$ supersymmetric
gauge theories}
The fact that we need a physical choice is natural
since entanglement itself is a physical property of the theory.
It should be kept in mind that an analogous choice is present for 
conventional geometric entanglement entropies \cite{TPS};
the notion of the geometric entanglement
depends on the physical choice of operationally
accessible interactions and measurements.

We can think of the choice 
of the compactification manifold $\mathcal{S}$
as a IR regulator of the theory.
The fact that theory-space entanglement depends on 
$\mathcal{S}$ is somewhat analogous to the fact that 
the conventional geometric entanglement entropy
depends on the choice of the spatial region $A$.

For continuous systems (such as quantum field theories) 
there are also UV divergences.
If we choose a small UV regulator $\epsilon$,
the leading contribution diverges as powers of $1/\epsilon$.
However, as in the case of geometric entropies, 
we expect that the subleading constant (i.e.\ order $\epsilon^0$) term or the 
coefficient of $\log\epsilon$ term is universal,
depending on whether the dimension is odd or even.
Note that this does not follow from the corresponding statement
for geometric entanglement entropies, since theory-space entanglement 
is different from geometric entanglement.

{\bf Comparison with Lattice Gauge Theories.}---
Some readers might be alarmed by the non-factorization 
of the Hilbert space \eqref{Hembed}, since
standard treatment of entanglement entropy 
assumes factorization. However, let us point out that
the factorization actually in general does not hold, even
for conventional geometric entanglement 
entropies (see Refs.~\cite{nonfactor}).

The issue arises for gauge theories.
For concreteness let us consider lattice gauge theories.
In the Hamiltonian formulation the gauge-invariant degrees of freedom
are given by strings of non-Abelian electric fluxes, and are 
not localized in space \cite{Hamiltonian}. 
Such fluxes in general spread both in regions $A$
and its complement $B$, and the spatial division 
violates the Gauss law on the boundary $\partial A$.
This explains the non-factorization of the Hilbert space.

To put it another way,
the problem is that in lattice gauge theories the basic degrees of freedom 
resides in the links connecting vertices,
and not in the vertices.
The boundary $\partial A$ pass through some of the links,
which are charged under some of the global symmetries.
Note that this is not just a conceptual problem, 
but is of practical importance
for numerical simulations of entanglement entropy.

\begin{figure}[htbp]
\centering{\includegraphics[scale=0.22]{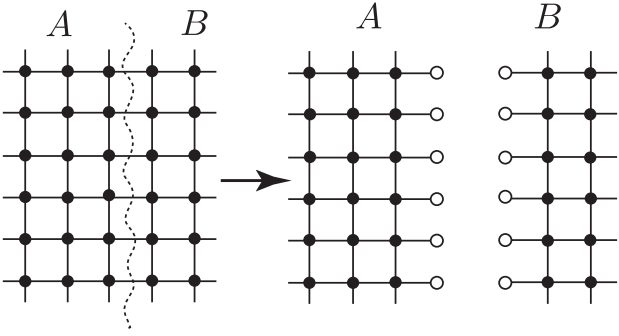}}
\caption{In lattice gauge theories, the definition of geometric
entanglement entropy requires the introduction of new vertices
(colored white)
and splitting of the link variables on the boundary.}
\label{fig.divide}
\end{figure}

To define geometric entropy for lattice gauge theories \cite{nonfactor},
we associate a new vertex for each link on the boundary
and divide the link into two smaller links, one associated with region $A$
and another region $B$. We then define the Hilbert space
$\mathcal{H}_{A}$ 
($\mathcal{H}_{B}$) to be the 
functionals of the connections of the links in region $A$ ($B$) 
which are gauge invariant with respect to the 
gauge transformations associated with the vertices in the interior of
 $A$ ($B$) but not necessarily with 
respect to the newly introduced vertices on the boundary (Fig.~\ref{fig.divide}).
We then have the natural embedding \eqref{Hembed}
and the geometric entanglement entropy is defined by 
\eqref{def3}, \eqref{def1}.
This is very analogous to the definition of the theory-space entanglement
above.

\begin{figure}[htbp]
\centering{\includegraphics[scale=0.24]{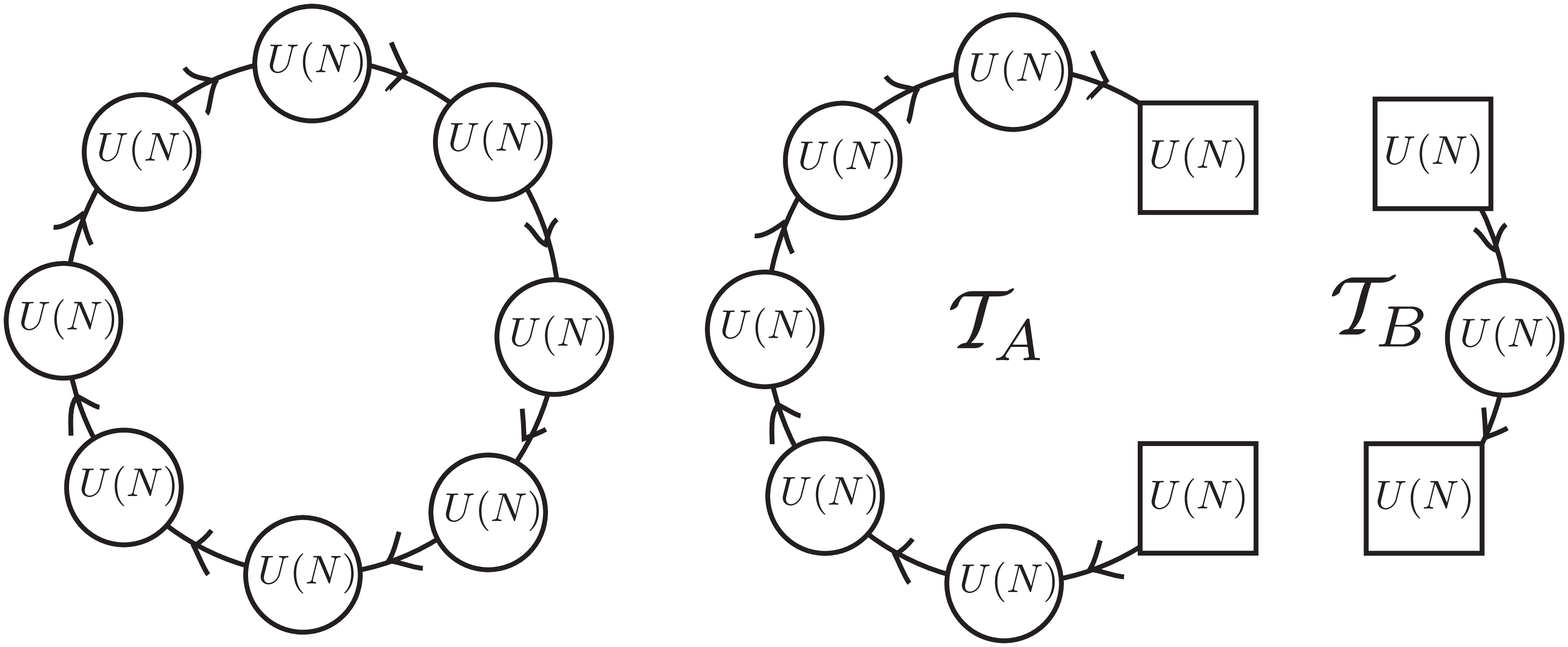}}
\caption{The quiver diagram (left) for the deconstruction of an extra dimension.
The circle direction along the quiver turns into a geometric extra
 dimension
in the IR. The theory-space entanglement defined by dividing the quiver into 
two (right) corresponds to a geometric entropy of the deconstructed theory.}
\label{fig.quiver}
\end{figure}

The analogy goes even further in the context of 
deconstruction \cite{deconstruction,deconstruction2}.
Let us begin with a quiver diagram on the circle, where quiver is simply a 
graph consisting of vertices and links (Fig.~\ref{fig.quiver}). 
Given a quiver we can construct a 
gauge theory by the rule that (1) we associate a $U(N_v)$ gauge group 
to each vertex $v$ and (2) we associate a bifundamental matter with respect to 
$U(N_v)\times U(N_w)$ for a link connecting vertices $v$ and $w$.
The precise matter content can vary depending on the context,
for example the amount of supersymmetry and the dimensionality of spacetime.
For example (in the original example of Ref.~\cite{deconstruction})
let us consider the 4d theory,
the link of the quiver is oriented,
and the associated matter is a Weyl fermion with chirality determined by the 
orientation. For simplicity we take $N_v$ to be independent of $v$,
and denote the corresponding integer by $N$.

Now the claim of Ref.~\cite{deconstruction,deconstruction2} is that in the IR limit and in
the limit of large number of quiver vertices,
the 4d theory on the Higgs branch
coincides with the 5d gauge theory, where only one of the 
directions is latticed
the direction of the quiver
becomes the extra dimension $S^1_{\rm extra}$ in the IR.

Since we have a 5d lattice gauge theory, 
we can define the geometric entanglement entropy along the 
$S^1_{\rm extra}$, by dividing $S^1_{\rm extra}$ into two.
More precisely 
let us take 5d lattice gauge theory dimensionally  
reduced on a compact 3-manifold $\mathcal{S}$, 
and consider the geometric entanglement for the 
resulting 2d theory on 
$\mathbb{R}_t\times S^1_{\rm extra}$. \footnote{Alternatively 
we could choose to latticize all the four dimensions.} 

As we have seen already, the definition of geometric entanglement 
in lattice gauge theories involves
introducing new vertices on the boundaries of regions $A, B$.
In the language of 4d quiver gauge theories, adding a node is translated
into adding $U(N)$ symmetry. Since the Hilbert spaces $\mathcal{H}_{A,B}$ are
not necessarily invariant under the $U(N)$ symmetry (as we discussed above),
we should regard the $U(N)$ as a global symmetry acting on 
$\mathcal{H}_{A,B}$;
to obtain $\mathcal{H}$ we need to gauge this symmetry. 
Since these are the same ingredients as
in the definition of theory-space entanglement above \footnote{
The symmetry gauged in this case is $\prod_v U(N_v)$, where $v$ runs over
all the links on the boundary.
}, 
we learn that the geometric entanglement in the deconstructed 5d theory 
(dimensionally reduced on $\mathcal{S}$) coincides with the 
theory-space entanglement for the 4d quiver gauge theory 
(defined on the same manifold 
$\mathcal{S}$)!
In other words we naturally arrive at the definition of the 
theory-space entanglement if 
we want to extend the notion of geometric entanglement of the 
deconstructed theory to 
the quiver gauge theory.
This is one justification for our definition, and illustrates nicely the 
close relation between geometric and theory-space entanglement.

{\bf Geometric/Theory-Space Duality}---
Let us provide another (and more non-trivial) example of the relation between 
geometric and theory-space entanglement.

This examples deals with the cause of 4d $\mathcal{N}=2$ superconformal field theories
arising from the compactification of 6d $(2,0)$ theories of type $A_N$ on 
a punctured Riemann surface $C$ \cite{Gaiotto}.
From the viewpoint of 4d gauge theory, the geometry $C$ 
is the defining data of the 4d theory.

\begin{figure}[htbp]
\centering{\includegraphics[scale=0.25]{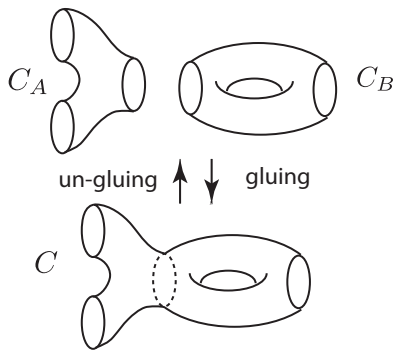}}
\caption{The pants decomposition of the Riemann surface.
In the geometric/theory-space duality of entanglement entropy,
the gluing/un-gluing here is translated into the
gauging/un-gauging of Fig.~\ref{Tglue}.
}
\label{fig.pants}
\end{figure}

A punctured Riemann surface $C$ can be decomposed into a collections of
 three-punctured spheres (trinions) (Fig.~\ref{fig.pants}). 
This is known as a pants decomposition.
In the theories defined in Ref.~\cite{Gaiotto}, 
a trinion is associated with a theory called $T_N$, with global 
symmetries $SU(N)^3$;
\footnote{Here we only consider the so-called full punctures. 
We can generalize the discussion to more general punctures
labeled by Young diagrams.}
each of the $SU(N)$ symmetries are associated with one of the punctures.
When we glue such trinions, we gauge the diagonal of the associated $SU(N)$ 
global symmetries; this is the gauging procedure of \eqref{Tfactor}.
In other words, gauging of \eqref{Tfactor} for 4d gauge theories
is translated into the geometrical operation of gluing
$C=C_A\cup C_B$.
Different choices of pants decompositions are argued to be 
different descriptions of the same 4d $\mathcal{N}=2$ superconformal
 IR fixed point, and thus are S-dual to each other.

Since the definition of the theory involves a gauging \eqref{Tfactor}, 
we can define the theory-space entanglement by
compactifying the 4d theory on $\mathbb{R}_t\times \mathcal{S}$,
where $\mathcal{S}$ is a compact 3-manifold.
The Hilbert space $\mathcal{H}_{\mathcal{T}[C]}$ for our 4d theory
$\mathcal{T}[C]$
depends on the choice of $\mathcal{S}$.
We here choose a
1-parameter family of the 3-sphere $S^3_b$ whose 
metric is given by $b^2(x_1^2+x_2)+b^{-2}(x_1^2+x_2)=1$ \cite{HHL}.

Now the surprise is that the Hilbert space 
$\mathcal{H}_{\mathcal{T}[C]}$
contains a subspace (``BPS Hilbert space'')
$\mathcal{H}_{\mathcal{T}[C]}^{\rm BPS}$
which coincides with the 
Hilbert space of the 
3d $SL(N)$ Chern-Simons theory
on the spatial Riemann surface:
\begin{equation}
\mathcal{H}[S^3_b]_{\mathcal{T}[C]}^{\textrm{4d BPS}}=\mathcal{H}^{\textrm{3d\,CS}}[{\mathcal{C}}]  \ ,
\label{HH}
\end{equation}
where the deformation parameter $b$ of $S^3_b$ is translated into the 
level $t$ of 3d $SL(N)$ Chern-Simons theory \cite{TY}.
This is part of the statement of the ``3d/3d duality'' \cite{TY} (see
also Refs.~\cite{3d3d, NW}).
In fact, 
we could regard $S^4_b$ of Refs.~\cite{Pestun,AGTW}
as $S^3_b$ fibered over an interval, with boundary conditions at both ends.

Let us consider the 6d $(2,0)$ theory on 
$\mathbb{R}_t\times S^3_b\times C$. We can regard this either as
(i) $(\mathbb{R}_t\times S^3_b) \times C$, giving rise to 
4d $\mathcal{N}=2$ theory on $\mathbb{R}_t\times S^3_b$
or (ii) $(\mathbb{R}_t\times C) \times S^3_b$, giving rise to a 
3d $SL(N)$ Chern-Simons theory on $\mathbb{R}_t\times C$ \cite{TY}.
Since \eqref{HH} is the equivalence of the Hilbert space,
we automatically have the equivalence of 
the density matrix and the corresponding 
entanglement entropies.
The correspondence is rather non-trivial since 
gauging of 4d $\mathcal{N}=2$ theories (Fig.~\ref{Tglue}) 
is translated into the
geometrical gluing operation on the 
2d surface (Fig.~\ref{fig.pants}); the 
theory-space entanglement in the 
BPS Hilbert space of the 4d theory \footnote{Due to boson-fermion
cancellation the partition function on $S^4_b$ is the 
same for the BPS
and the full Hilbert spaces \cite{Pestun}. 
However it is not clear if their theory-space
entanglements are the same.} is identified with the 
geometric entanglement in the 
3d $SL(N)$ Chern-Simons theory on the geometric surface $C$!

We can also discuss a similar correspondence 
for a different compactification manifold
 $\mathcal{S}$; 
we can for example take the 4d theory on $S^1\times S^3$, but with a 
twist along the $S^1$ direction \cite{SCI}. 
The corresponding 3d theory is the $SU(N)$ Chern-Simons theory on 
$S^1\times C$, which in turn gives 2d $q$-deformed 
Yang-Mills theory \cite{AOSV} on $C$ (cf. Ref.~\cite{Fradkin}).
\footnote{It is tempting to speculate that similar correspondence 
holds for more general theories, such as those in Refs.~\cite{YBE},
which discuss the relations between 4d superconformal indices and 
2d spin chains.
}

Note in both of these cases 
the definition of entanglement entropy depends on the choice of the 
duality frame, and is not S-duality invariant.

{\bf Strong Subadditivity.}--- 
Geometric entropies satisfy one crucial relation, the strong subadditivity
\begin{equation}
\begin{split}
&S_{\rm ent}(A_1\cup A_2) +S_{\rm ent}(A_2\cup A_3) \\
& \qquad \ge S_{\rm ent}(A_1\cup A_2 \cup A_3) +S_{\rm ent}(A_2) \ ,
\label{subadd}
\end{split}
\end{equation}
for three spatial regions $A_{1,2,3}$. 

We conjecture that there exists a counterpart for this statement
in theory-space entanglement.
To be concrete, suppose that the theory $\mathcal{T}$ has a decomposition into four:
\begin{equation}
\mathcal{T}=\mathcal{T}_{1} \cup_{G_1} \mathcal{T}_{2} \cup_{G_2} \mathcal{T}_3 \cup_{G_3} \mathcal{T}_4 \ .
\label{Ttotal}
\end{equation}
It is then natural to define 
\begin{equation}
\begin{split}
&\mathcal{T}_{1\cup 2}=\mathcal{T}_1\cup_{G_1} \mathcal{T}_2  \ ,
\mathcal{T}_{2\cup 3}=\mathcal{T}_2\cup_{G_2} \mathcal{T}_3  \ , \\
&\mathcal{T}_{1\cup 2\cup 3}=\mathcal{T}_1\cup_{G_1} \mathcal{T}_2 \cup_{G_1} \mathcal{T}_3  \ .
\end{split}
\end{equation}
The counterpart of \eqref{subadd} is 
\begin{equation}
S_{\rm th}(\mathcal{T}_{1\cup 2}) +S_{\rm th}(\mathcal{T}_{2\cup 3}) \ge S_{\rm th}(\mathcal{T}_{1\cup 2 \cup 3}) +S_{\rm th}(\mathcal{T}_{2}) \ ,
\label{subadd2}
\end{equation}
where $S_{\rm th}$ is the theory-space entanglement defined with respect to the total theory $\mathcal{T}$ 
in \eqref{Ttotal}. The proof of \eqref{subadd2} will be similar to that of \eqref{subadd} (see e.g. Ref.~\cite{Nielsen}),
however we have to carefully take the non-factorization \eqref{Hembed} into account.

{\bf Summary and Discussion.}--- In this paper we discussed a new notion
of entanglement,
the {\it theory-space enanglement}, 
which quantifies the entanglement of two theories interacting through
gauge interactions. 

While the idea might sound unfamiliar at first, the definition follows
that of the conventional geometric entropies, with the only difference
being that the division into two regions refers not to division 
in the geometric space, but in the more abstract theory space. 
Moreover we have shown that the theory-space
entanglement entropies for a class of theories
are equivalent to the geometric entanglement entropies 
for the dual theories.

The notion of the theory-entanglement entropies can further be generalized
--- it is not crucial for our definition that the interactions
between theories $A$ and $B$ are mediated by gauge interactions.
For example, 
the two theories can interact through Yukawa interactions
between bosons in $A$ and fermions in $B$.
This example is simpler than the case with gauge interactions since
there is no counterpart of dynamical gauge bosons.

Our theory-space entanglement entropies, when generalized in this way,
are a rather genral quantitative tool to measure entanglement between two 
theories interacting through some
(e.g.\ gauge)
long-range interactions.
There are many such examples in physics, 
indicating the utility of the entanglement
in a wide range of physical phenomena. 
Let us here mention a few of them for illustration.

A good example for our entanglement is the discussion of vacuum entanglement
 in Refs. \cite{vacuum},
which analyzes the vacuum entanglement for a scalar field
interacting with two atoms/detectors $A$ and $B$ --- in this case the 
``theories'' are simply atoms/detectors and their mutual interactions
are mediated by the scalar field. We can also generalize the discussion
 there by replacing the scalar field by the gauge field.

Another example is the gauge-mediated supersymmetry breaking scenario
\cite{GMSB} discussed previously in this Letter. In this case the
theory-space entanglement quantifies the degree to which the 
standard mode physics (or its grand unified versions) is sensitive to
the physics in the hidden sector. This partly answers the question of
whether we can distinguish between different models of hidden sectors,
which is of great phenomenological interest.

Interestingly, in many cases the separation in the abstract theory space
actually coincides with the physical separation in the spatial regions,
since the theory $A, B$ could for example refer to materials
placed/localized in some geometrical regions, and their interactions are given by
long-range foreces such a photons.
Such geometrical separation also occurs when we choose to take $A, B$ to be
theories on the branes in the brane-world scenario, where the two
theories are localized in extra dimensions
and interact though gravity.

More ambitiously, we believe that theory-space entanglement 
will provide useful tools to explore the {\it space of 
quantum field theories} in various dimensions,
and learn about their mutual relations,
perhaps along the lines of the Zamolodchikov metric for CFT.
A general inequality among theory-space entanglement entropies,
such as the strong subadditivity discussed in this Letter,
could constrain the possible forms of interactions
between the two theories $A$ and $B$.

Finally, it would be interesting to systematically compute
the theory-space entanglement entropies for concrete examples.
The replica trick \cite{replica}, which works well for geometric entanglement
entropies, in itself does not work here since the 
the meaning of the $n$-fold cover in the theory space is not clear.
We can instead choose to compute the theory-space entanglement 
order by order in the gauge coupling constant 
in the perturbative expansion.
It would also be interesting to ask if theory-space entanglement has 
the counterpart of the Ryu-Takayanagi formula in the holographic description.
 
\acknowledgments

The author would like to thank Aspen Center for Physics 
(NSF Grant No.\ 1066293) for 
hospitality, and would like to thank 
A.~Gorsky, I.~Klebanov, J.~Maldacena, T.~Nishioka and Y.~Tachikawa
for comments and discussions.

\end{document}